\documentstyle[12pt]{article}

\begin{document}
\thispagestyle{empty}

\begin{center}
\LARGE \tt \bf{Helical ${\alpha}$-dynamos as twisted magnetic flux
tubes in Riemannian space}
\end{center}

\vspace{1.5cm}

\begin{center} {\large L.C. Garcia de Andrade \footnote{Departamento de
F\'{\i}sica Te\'{o}rica - Instituto de F\'{\i}sica - UERJ

Rua S\~{a}o Fco. Xavier 524, Rio de Janeiro, RJ

Maracan\~{a}, CEP:20550-003 , Brasil.E-mail:garcia@dft.if.uerj.br}}
\end{center}

\vspace{2.0cm}

\begin{abstract}
Analytical solution of ${\alpha}$-dynamo equation representing
strongly torsioned helical dynamo is obtained in the thin twisted
Riemannian flux tubes approximation. The $\alpha$ factor possesses a
fundamental contribution from torsion which is however weaken in the
thin tubes approximation. It is shown that assuming that the
poloidal component of the magnetic field is in principle
time-independent, the toroidal magnetic field component grows very
fast in time, actually it possesses a linear time dependence, while
the poloidal component grows under the influence of torsion or twist
of the flux tube. The toroidal component decays spatially with as
$r^{-2}$ while vorticity may decay as $r^{-5}$ (poloidal component)
 where r represents the radial distance from the magnetic axis of
flux tube. Toroidal component of vorticity decays as $r^{-1}$. In
turbulent dynamos unbounded magnetic fields may decay at least as
$r^{-3}$. \vspace{0.5cm} \noindent {\bf PACS
numbers:\hfill\parbox[t]{13.5cm}{02.40.Hw-Riemann geometries}}
\end{abstract}
\newpage
\section{Introduction}
 Despite of the success of the application of the numerical simulations to the dynamo problem \cite{1}
 in plasma astrophysics \cite{2} and in the stretch-twist-fold (STF)
 Vainshtein-Zeldovich \cite{2} mechanism, recently new dynamo
 analytical solutions have been found \cite{3} by using the
 conformal mapping technique in Riemannian manifolds from old Arnold
 cat dynamo metric \cite{2}. Earlier T. Kambe \cite{4} found
 simultaneous vortical and magnetohydrodynamic (MHD) solutions. In this paper an helical dynamo \cite{8} solution of self-induction
 is obtained in vortical strongly torsioned thin twisted magnetic
 flux tubes in Riemannian space \cite{5} where the MHD equations are linear in the magnetic field
 and nonlinear in the velocity flow. Assuming that the poloidal is time-independent the toroidal component of the magnetic field
 grows fast in time, actually it grows linear and not exponential. The rate of growing of the toroidal component depending
 on the inverse squared of the radial distance of the magnetic axis which possesses curvature and torsion. Recently, Hanasz and Lesch \cite{6} have used also a conformal Riemannian metric in
 ${\cal{E}}^{3}$ to the galactic dynamo magnetic flux tubes. Pioneering work on the magnetic flux tubes as dynamos was done
 earlier by Schussler \cite{7} , however in his work tubes were untwisted and straight. The main advantage of the
 investigation of the isolated flux tube
 dynamo is that one is able to investigate the curvature and twist contributions of the tube to the dynamo action.
 Twist is actually related to the torsion of the magnetic axis of the tube, which makes the words strong torsion equivalent to
 strong twist, which physically is important to the twist-kink relation investigated by Alfven \cite{8}. Helical dynamo here is understood as the one
 where the flow describes a circular helix where torsion and curvature are constants and equal. The paper is organized as follows: In section
 II the dynamo solution in the Riemann metric representing flux rope (twisted tubes) is obtained. In section III the
 approximate solution is presented. In section IV conclusions
 are given.
 \section{Helical dynamos in Riemannian space}
In this section we shall be concerned with solving the MHD equations
in the curved coordinates of a thin twisted magnetic flux tube of
Riemann metric
\begin{equation}
ds^{2}=dr^{2}+r^{2}d{{\theta}_{R}}^{2}+{K^{2}}(s)ds^{2} \label{1}
\end{equation}
which represents a Riemannian line element
\begin{equation}
ds^{2}=g_{ij}dx^{i}dx^{j} \label{2}
\end{equation}
 if the tube coordinates are $(r,{\theta}_{R},s)$ \cite{1} where
 ${\theta}(s)={\theta}_{R}-\int{{\tau}ds}$ and $\tau$ is the Frenet
 torsion of the tube axis, $K(s)$ is given by
\begin{equation}
{K^{2}}(s)=[1-r{\kappa}(s)cos{\theta}(s)]^{2} \label{3}
\end{equation}
Since we are considered thin magnetic flux tubes, this expression
shall be taken as $K\approx{1}$ in future computations. In
curvilinear coordinates the Riemannian Laplacian operator
${\nabla}^{2}$ \cite{1} is
\begin{equation}
{\nabla}^{2}=\frac{1}{\sqrt{g}}{\partial}_{i}[\sqrt{g}g^{ij}{\partial}_{j}]
\label{4}
\end{equation}
where ${\partial}_{j}:=\frac{{\partial}}{{\partial}x^{j}}$ and
$g:=det{g_{ij}}$ where $g_{ij}$ are the covariant components of the
Riemann metric of flux rope.Let us now start by considering the MHD
field equations
\begin{equation}
{\nabla}.\vec{B}=0 \label{5}
\end{equation}
\begin{equation}
\frac{{\partial}}{{\partial}t}\vec{B}={\nabla}{\times}[{\alpha}\vec{B}]={\alpha}{\nabla}{\times}\vec{B}+{\nabla}{\alpha}{\times}{\vec{B}}
 \label{6}
\end{equation}
called the ${\alpha}$-dynamo equation \cite{13}. Sometimes the
${\alpha}:=<\vec{v}.\vec{\omega}>$ parameter is constant but here we
shall be considering the more general case where it depends on the
radial and poloidal coordinate. Here
${\omega}:={\nabla}{\times}\vec{v}$ is the vorticity of the dynamo
flow. Equation (\ref{3}) represents the self-induction equation. The
vectors $\vec{t}$ and $\vec{n}$ along with binormal vector $\vec{b}$
form the Frenet holonomic frame, which obeys the Frenet-Serret
equations
\begin{equation}
\vec{t}'=\kappa\vec{n} \label{7}
\end{equation}
\begin{equation}
\vec{n}'=-\kappa\vec{t}+ {\tau}\vec{b} \label{8}
\end{equation}
\begin{equation}
\vec{b}'=-{\tau}\vec{n} \label{9}
\end{equation}
where the dash represents the ordinary derivation with respect to
coordinate s, and $\kappa(s,t)$ is the curvature of the curve, where
$\kappa=R^{-1}$. Here ${\tau}$ represents the Frenet torsion. The
gradient operator is
\begin{equation}
{\nabla}=\vec{t}\frac{\partial}{{\partial}s}+\vec{e_{\theta}}\frac{1}{r}\frac{\partial}{{\partial}{\theta}}+
\vec{e_{r}}\frac{\partial}{{\partial}r} \label{10}
\end{equation}
 Now we shall consider the analytical solution of the self-induction magnetic equation which represents a
 non-dynamo thin magnetic flux tube. Before the derivation of this result is obtained, we would like to point it out
 that it is not trivial, since the Zeldovich antidynamo theorem states that the two
 dimensional magnetic fields do not support dynamo action. Here, as is shown bellow, the flux tube axis
 possesses not only Frenet curvature, but torsion as well, and this
 last one vanishes in planar curves. The magnetic field does not
 possess a radial component and the magnetic field can be split inti its toroidal and poloidal components as
\begin{equation}
\vec{B}(r,s,t)={B_{\theta}}(t,r,{\theta}(s))+
B_{s}(r)\vec{t}\label{11}
\end{equation}
Now let us substitute the definition of the poloidal plus toroidal
magnetic fields into the self-induction equation,along with
expressions
\begin{equation}
\vec{e_{\theta}}=-\vec{n}sin{\theta}+\vec{b}cos{\theta} \label{12}
\end{equation}
and
\begin{equation}
\vec{e_{r}}=\vec{n}cos{\theta}+\vec{b}sin{\theta} \label{13}
\end{equation}
\begin{equation}
{\partial}_{t}\vec{e_{\theta}}={\omega}_{\theta}\vec{e}_{r}-{\partial}_{t}\vec{n}sin{\theta}+{\partial}_{t}\vec{b}cos{\theta}
\label{14}
\end{equation}
Considering the equations for the time derivative of the Frenet
frame given by the hydrodynamical absolute derivative
\begin{equation}
\dot{\vec{X}}={\partial}_{t}\vec{X}+[\vec{v}.{\nabla}]\vec{X}
\label{15}
\end{equation}
where $\vec{X}=(\vec{t},\vec{n},\vec{b})$ is used into the
expressions for the total derivative of each Frenet frame vectors
\begin{equation}
\dot{\vec{t}}={\partial}_{t}\vec{t}+[{\kappa}'\vec{b}-{\kappa}{\tau}\vec{n}]
\label{16}
\end{equation}
\begin{equation}
\dot{\vec{n}}={\kappa}\tau\vec{t} \label{17}
\end{equation}
\begin{equation}
\dot{\vec{b}}=-{\kappa}' \vec{t} \label{18}
\end{equation}
therefore leading to the following values of respective partial
derivatives of the Frenet frame
\begin{equation}
{\partial}_{t}\vec{t}=-{\tau}{\kappa}[1-{\kappa}{\tau}^{-2}\frac{v_{\theta}}{r}]\vec{n}
\label{19}
\end{equation}
\begin{equation}
{\partial}_{t}{\vec{n}}={\tau}{\kappa}[1-{\kappa}\vec{\tau}^{-2}\frac{v_{\theta}}{r}]\vec{t}+\frac{v_{\theta}}{r}\vec{b}
\label{20}
\end{equation}
\begin{equation}
{\partial}_{t}{\vec{b}}={\kappa}{\tau}^{-1}\frac{v_{\theta}}{r}\vec{n}
\label{21}
\end{equation}
where use has been made of the hypothesis that $\dot{\vec{b}}=0$ or
${\kappa}'(t,s)=0$, which means that the curvature only depends on
time. Substitution of these vectorial expressions into expression
(\ref{14}) yields
\begin{equation}
{\partial}_{t}\vec{e_{\theta}}=
-{\omega}_{\theta}{\vec{e}}_{r}+{\gamma}[{\tau}_{0}sin{\theta}(1-{\tau}_{0})\vec{t}+{\tau}_{0}cos{\theta}\vec{n}
-{\tau}_{0}sin{\theta}\vec{b}] \label{22}
\end{equation}
where ${\gamma}:=(v_{s}-\frac{1}{r}{{\tau}_{0}}^{-1}v_{\theta})$.
The other dynamical equation for the Frenet holonomic frame is
\begin{equation}
{\partial}_{t}\vec{e_{\theta}}=[{\tau}_{0}sin{\theta}\vec{t}-[{\theta}_{\theta}+{\tau}_{0}]cos{\theta}\vec{n}
- [{\omega}_{\theta}+{\tau}_{0}]sin{\theta}\vec{b}] \label{23}
\end{equation}
note that in the mean field dynamo case, where
$\vec{v}=\vec{v}(\vec{B})$ , equation (\ref{6}) is an eigenvalue
problem equation. Dynamo operators and eigenvalue of dynamos in
compact Riemannian manifolds have been previously investigated by
Chicone and Latushkin \cite{9}. Substitution of previous equations
into equation (\ref{6}) and splitting these equations along the
components of the Frenet frame $(\vec{t},\vec{n},\vec{b})$ yields
the following three scalar equations
\begin{equation}
{\partial}_{t}{B_{s}}+[sin{\theta}{\gamma}-{\tau}_{0}]{\tau}_{0}B_{\theta}={\partial}_{r}({\alpha}B_{\theta})
\label{24}
\end{equation}
\begin{equation}
-{\partial}_{t}B_{\theta}sin{\theta}-B_{\theta}{\omega}_{\theta}cos{\theta}+(cos{\theta}B_{\theta}{\gamma}-{\tau}_{0}B_{s})
{\tau}_{0}= A cos{\theta}-C sin{\theta} \label{25}
\end{equation}
\begin{equation}
{\partial}_{t}B_{\theta}cos{\theta}-B_{\theta}({\omega}_{\theta}+{\tau}_{0})sin{\theta}
=A sin{\theta}+C cos{\theta} \label{26}
\end{equation}
where functions A and C are given respectively by
\begin{equation}
A:=
\frac{1}{r}({\partial}_{\theta}{\alpha})B_{s}-({\partial}_{s}{\alpha})B_{\theta}-{\alpha}{\partial}_{s}B_{\theta}
\label{27}
\end{equation}
and
\begin{equation}
C:=({\partial}_{r}{\alpha})B_{s}+\frac{{\alpha}}{r}{\partial}_{r}B_{r}
\label{28}
\end{equation}
In the next section we shall solve find an approximate solution for
strong torsioned ${\alpha}-dynamos$ tubes. Helical dynamo hypothesis
of ${\kappa}_{0}=constant={\tau}_{0}$ has been taken throughout
these computations.
\section{Analytical solution of helical dynamos in flux tubes}
From the solenoidal equation for the magnetic field one obtains
\begin{equation}
{\partial}_{s}B_{\theta}=B_{\theta}{{\tau}_{0}}^{2}r sin{\theta}
\label{29}
\end{equation}
one obtains the value for the poloidal component as
\begin{equation}
B_{\theta}=B_{0}exp({\tau}_{0}r cos{\theta}) \label{30}
\end{equation}
with this expression and the expressions for the value of ${\alpha}$
which can be obtained as
\begin{equation}
{\alpha}=v_{\theta}{\omega}_{\theta}+v_{s}{\omega}_{s} \label{31}
\end{equation}
To compute this important factor ${\alpha}$ one computes the
vorticity ${\vec{\omega}}$ as
\begin{equation}
{\omega}_{r}=-{\partial}_{s}v_{\theta}={\kappa}_{0}{\tau}_{0}rv_{\theta}sin{\theta}
\label{32}
\end{equation}
where we have used in this equation the physical assumption of the
incompressibility of the dynamo flow. The remaining vorticity
expressions are
\begin{equation}
{\omega}_{\theta}=-{\partial}_{r}v_{s} \label{33}
\end{equation}
\begin{equation}
{\omega}_{s}={\partial}_{r}v_{\theta}+\frac{1}{r}v_{\theta}
\label{34}
\end{equation}
This allows us to obtain the following value for ${\alpha}$
\begin{equation}
{\alpha}=\frac{1}{r}v_{0}exp(-r{\tau}_{0}cos{\theta})
 \label{35}
\end{equation}
By making use of thin tube approximation where we consider we are
close to the magnetic flux tube axis $(r=0)$, one may use the
following approximation
\begin{equation}
{\alpha}=\frac{1}{r}v_{0} \label{36}
\end{equation}
which yields
\begin{equation}
{\partial}_{s}{\alpha}=-\frac{1}{r^{2}}v_{0} \label{37}
\end{equation}
which upon substitution in equation (ref{1}) yields
\begin{equation}
{\partial}_{t}{B_{s}}=[{{\tau}_{0}}^{2}-\frac{v_{0}}{r^{2}}]B_{\theta}
\label{38}
\end{equation}
\begin{equation}
{\partial}_{t}{B_{s}}=-\frac{v_{0}}{r^{2}}B_{0} \label{39}
\end{equation}
which simply solves to
\begin{equation}
{B_{s}}=-\frac{v_{0}}{r^{2}}B_{0}t \label{40}
\end{equation}
which shows that the toroidal field grows linearly in time which
close to the magnetic axis may represent a very fast dynamo.
Comparison between the poloidal and toroidal field one obtains
\begin{equation}
\frac{B_{\theta}}{B_{s}}{\approx}\frac{{v_{0}}}{r^{2}}t \label{41}
\end{equation}
In regions not so close to the magnetic axis where torsion may
dominate one obtains an alternative solution in which however the
magnetic toroidal field still grows in time as
\begin{equation}
{B_{s}}{\approx}{\tau_{0}}^{2}t \label{42}
\end{equation}
Substitution of these equations in the remaining dynamo equation
yields the poloidal component of vorticity as
\begin{equation}
{\omega}_{\theta}=\frac{v_{0}t}{r^{5}}\label{43}
\end{equation}
which implies that
\begin{equation}
{v}_{s}=\frac{v_{0}t}{4r^{4}}\label{44}
\end{equation}
and the other vorticity components are
\begin{equation}
{\omega}_{r}{\approx}
-{{\tau}_{0}}^{2}rv_{0}exp({\tau}_{0}rcos{\theta}) \label{45}
\end{equation}
and ${\omega}_{r}=0$ at the magnetic axis. Finally
\begin{equation}
{{\omega}_{s}}=\frac{v_{0}}{r}exp({\tau}_{0}rcos{\theta}\label{46}
\end{equation}
which decays slower as we go away from the magnetic axis of flux
tube dynamo.
\section{Conclusions}
In conclusion, we have used an approximate method of strong torsion
to find near the magnetic axis of the dynamo flux tube to analytical
(non-numerical) solutions of ${\alpha}$-dynamo equation. Previously
Ruzmaikin et al \cite{10} have found solutions of turbulent dynamos
where unbounded magnetic field could decay at least as $r^{-3}$
which is distinct to the bounded case of twisted magnetic flux tube
helical dynamo we have found here. The general equations of the
helical dynamo found in section II could be used to find out more
general solutions, with less degree of approximation that we used
here, for example by letting the magnetic poloidal component vary
with time and also considered the case of untwisted tubes. This can
be done elsewhere.
\section*{Acknowledgements}
I would like to dedicate this paper to the memory of Professor
Vladimir Tsypin, friend and teacher, a great physicist and
mathematician of plasmas, which taught me a great deal of the
applications of the Riemannian geometry to plasma physics. on the
occasion of his senventh birthday. I would like also to thank CNPq
(Brazil) and Universidade do Estado do Rio de Janeiro for financial
supports.


\newpage
\begin{thebibliography}{10}

\bibitem{1} S.I. Vainshtein, R. Sagdeev, and R. Rosner, Phys Rev E
56, 2 1605 (1997).
\bibitem{2} Ya B. Zeldovich, A.A. Ruzmaikin and D.D. Sokoloff, The
Almighty Chance, World sci. Press, Singapore (1990).  V. Arnold and
B. Khesin, Topological Methods in Hydrodynamics, Applied Mathematics
Sciences 125, Springer, Berlim (1998). V. Arnold, Ya B. Zeldovich,
A. Ruzmaikin and D.D. Sokoloff, JETP 81 (1981),n. 6, 2052. V.
Arnold, Ya B. Zeldovich, A. Ruzmaikin and D.D. Sokoloff, Doklady
Akad. Nauka SSSR 266 (1982) n6, 1357.
\bibitem{3} L.C. Garcia de Andrade, Stretch fast dynamo
mechanism in conformal Riemannian manifolds  14, october
issue,(2007).
\bibitem{4} T. Kambe, Geometrical theory of dynamical systems and fluid flows, World
scientific, Singapore (2000).
\bibitem{5} R. Ricca, Solar Physics 172
(1997),241.
\bibitem{6} M. Hanasz, H. Lesch,Astronomy and Ap 321,
(2007) 1007.
\bibitem{7} M. Schussler, Nature 288,150 (1980).
\bibitem{8} H. Alfven, Tellus 2, 74 (1950).
\bibitem{9} C. Chicone and Latushkin, Evolution Semigroups AMS-(1999).
\bibitem{10} A.A. Ruzmaikin, A. M. Shukurov and D. D. Sokoloff, Magnetic fields of Galaxies, Kluwer publishers (1988)\end{thebibliography}
\end{document}